\documentclass[cits]{PoS} 

\title{Ion-Channeling in Direct Dark Matter Crystalline Detectors }

\ShortTitle{Ion-Channeling in Direct Dark Matter Crystalline Detectors }

\author{\speaker{Graciela GELMINI}%
         \thanks{Talk  at IDM2010 (Identification of Dark Matter, 26-30 Jul. 2010, Montpellier, France) based on work done with Nassim Bozorgnia and Paolo Gondolo.}\\
        Dept. of Physics and Astronomy, University of California-Los Angeles\\
        E-mail: \email{gelmini@physics.ucla.edu}}

\abstract{The channeling of the recoiling nucleus in crystalline detectors after a WIMP collision would produce a larger scintillation or ionization signal in direct detection experiments than otherwise expected.  I present estimates of the importance of this effect in NaI, Si and Ge crystals,  using analytic models developed from the 1960's onwards to describe  channeling and blocking effects.}

\FullConference{Identification of Dark Matter 2010\\
                 July 26 - 30 2010\\
                 University of Montpellier 2, Montpellier, France}

\begin{document}

Channeling and blocking in crystals refer to the orientation dependence of ion penetration in crystals. ``Channeling" occurs when ions incident upon a crystal along symmetry axes and planes suffer a series of small-angle scatterings  that maintain them in the open ``channels"  in between rows or planes of lattice atoms and thus penetrate much further than non-channeled ions. Channeled ions  do not get close to lattice sites, where they would suffer a large-angle scattering which would take them out of the channel. ``Blocking" is the reduction  of the flux of ions originating in lattice sites along symmetry axes and planes due to the shadowing effect of the lattice atoms directly in front of the emitting lattice site (see e.g. the review by D. Gemmell~\cite{Gemmell:1974ub} and references therein).

Channeling and blocking effects in crystals are extensively  used in crystallography, 
  in measurements of short nuclear lifetimes, in the production of polarized beams etc. Channeling must be avoided in the implantation of   B, P and As atoms  in Si crystals to make circuits.   The channeling effect in NaI(Tl) crystals was first observed by Altman et al.~\cite{altman} in 1973 who observed that channelled ions produce more scintillation light because they loose practically all their energy via electronic stopping rather than nuclear stopping. 
When ions recoiling after a collision with a WIMP move along crystal axes and planes, they give all their energy to electrons, so the quenching factor $Q$ (the fraction of the energy deposited in a collision that goes into scintillation or ionization) is $Q \simeq1$ instead of  e.g. $Q_I \simeq 0.09$ and $Q_{Na} \simeq 0.3$ in NaI(Tl). The  potential importance of this effect for direct  dark matter detection was first pointed out  for NaI(Tl) by Drobyshevski  in 2007 and soon after by the DAMA collaboration~\cite{DAMA}. This last paper  gave an estimate of the channeling fraction of recoiling ions as function of the energy, in which the fraction  grew with decreasing energy to be $\simeq 1$ close to 1 keV. With this channeling fraction, the region of light WIMPs compatible with producing an annual modulation as measured by the DAMA collaboration shifted considerably to lower cross sections, by about one order of magnitude (see e.g. \cite{Savage-2009}).
Channeling could also produce a novel dark matter signature. Since the WIMP wind comes preferentially from the direction fixed to the galaxy towards which the Sun moves, Earth's daily rotation makes this direction change with respect to the crystal, which could produce a daily modulation in the measured  energy (equivalent to a modulation of the quenching factor). As first pointed out by Avignone, Creswick, Nussinov in 2008~\cite{Avignone}, this modulation depends on the orientation of the crystal with respect to the galaxy, thus it would be a background free dark matter signature.

Our calculations~\cite{BGG}
 of the fraction  of recoils that are channeled as function of recoil energy and direction use classical analytic models which started to be developed in the 60's and 70's, soon after channeling was discovered,   for  ions of energy MeV and higher. We use in particular Lindhard's model~\cite{Lindhard:1965}, supplemented by the planar channel models of Morgan and Van Vliet~\cite{Morgan} and the 1995-1996 work of G. Hobler~\cite{Hobler} on low energy channeling (applied to ion implantation in Si). In these models the rows and planes of lattice atoms are replaced by continuum strings and planes  in which the screened Thomas-Fermi potential  is averaged over a direction
parallel to a row or plane of lattice atoms to find the continuous potential $U$ (see Fig. 1.a for examples of $U$). The appearance of continuous  strings or planes can be understood as the overlap of  the ``Coulomb shadows" of individual atoms in a lattice row or plane behind the direction of arrival of a parallel beam of positive ions, when the beam arrives at a small enough angle, smaller than a critical angle $\psi_c$. Then, the individual shadows overlap forming a string or plane of thickness $\rho_c$ within which the scattered ions do not penetrate (see e.g. Fig. 4.6 of Ref.~\cite{ion-book}).  We included just one string or plane, which is a good approximation except  when $\rho_c$ approaches the radius of the channel. 

Lindhard proved that for channeled ions the ``transverse energy" $E_{\rm perp}=  M {\rm v}_{\rm perp}^2/2 + U  \simeq E \phi^2 + U$ is conserved, where $M, ~E=M {\rm v}^2/2$ are the mass and kinetic energy of the propagating ion, ${\rm v}_{\rm perp} = {\rm v} \sin\phi \simeq {\rm v} \phi$ is the component of the velocity perpendicular to the string or plane and $U$ is the continuum axial or planar potential at the position of the ion.
Thus, for a particular  $E_{\rm perp}$ (determined by the recoil energy $E$, initial recoil angle $\phi_i$ and initial potential  and $U_i$),  a channeled propagating ion does  not approach the string or plane closer than a minimum distance $\rho_{\rm min}$ (for which ${\rm v}_{\rm perp} =0$) and far away from the string or plane, close to the middle of the channel, the ion moves on a trajectory forming an angle $\psi$ with the string or plane, given by $E_{\rm perp} = E \phi_i^2 + U_i= U(\rho_{\rm min})= E \psi^2  +U_{\rm middle}$. Channeling occurs when $\rho_{\rm min} > \rho_c$, which amounts to $ \psi=  \sqrt{\left[ U(\rho_{\rm min})- U_{\rm middle})\right]/ E} \leq \psi_c = \sqrt{\left[ U(\rho_c)- U_{\rm middle})\right]/ E} $. 

\begin{figure} 
\includegraphics[width=.51\textwidth]{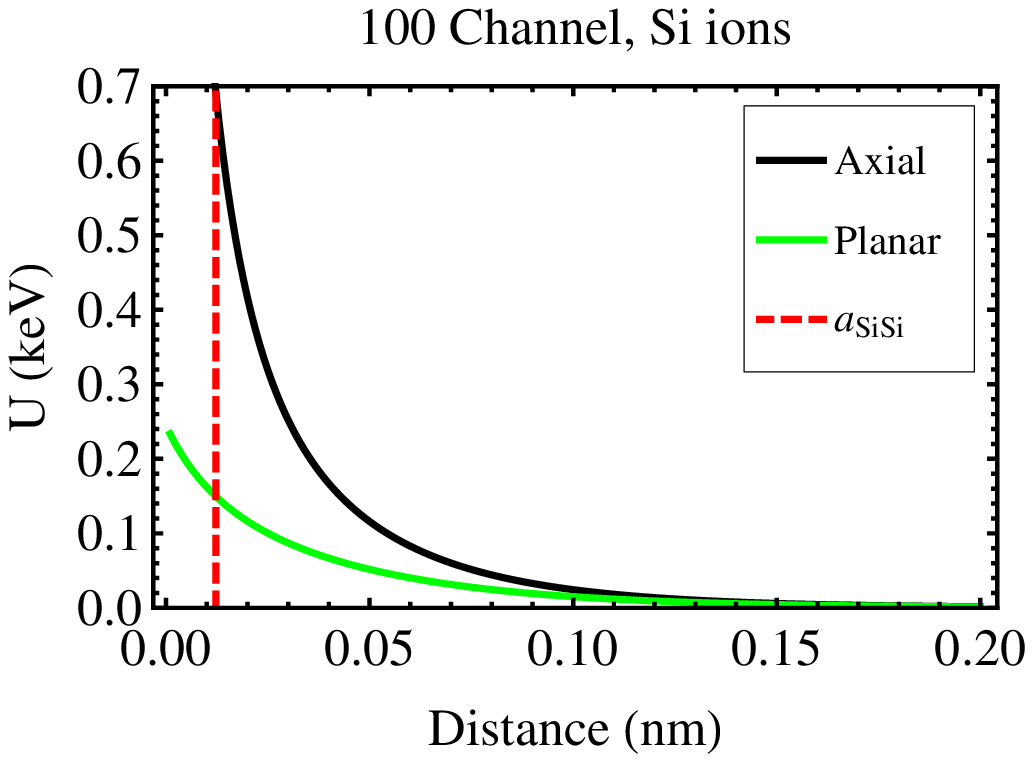} 
\includegraphics[width=.41\textwidth]{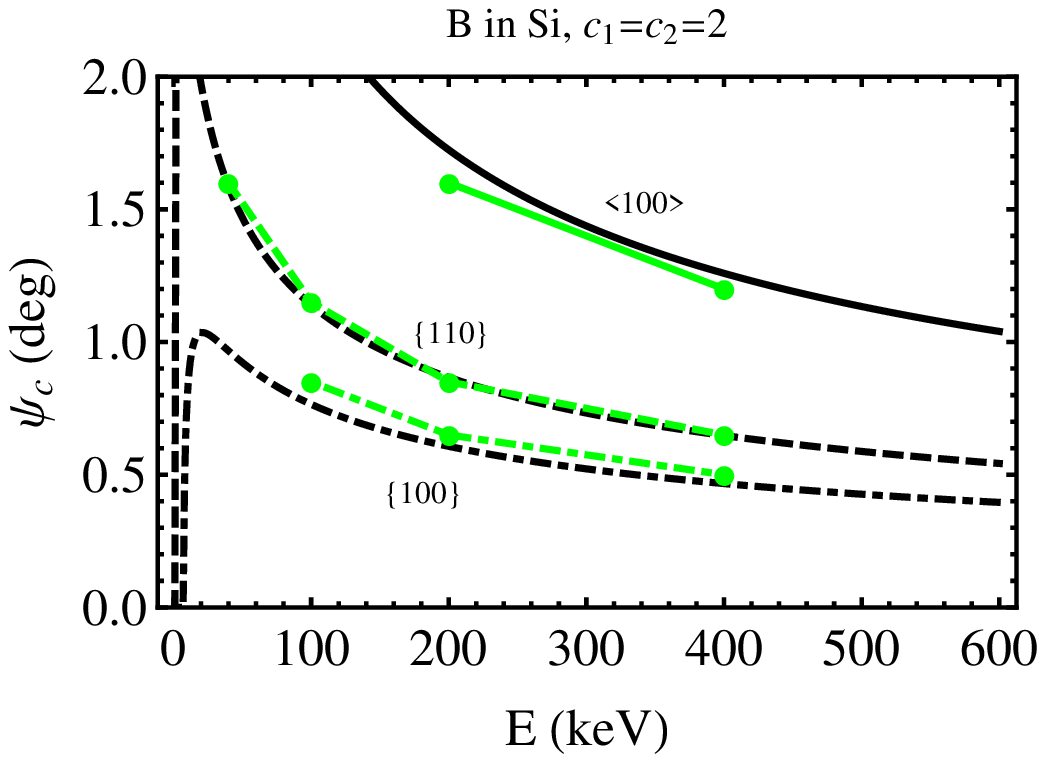} 
\caption{a. (left) Continuum potential for a string (black) and plane (green) for a Si ion propagating in Si as function of the distance perpendicular to the string or plane. b. (right) Oberved (green) and predicted with $c=2$ (black) channeling angles for B ions propagating in Si at room T.} 
\label{fig1} 
\end{figure} 

All the difficulty of this approach resides in calculating $\rho_c$.  Including temperature ($T$) effects, the critical distance can be approximated by  $\rho_c (E,T) = \sqrt{\rho_c^2(E) +  \left[c~u_1(T)\right]^2}$, where $\rho_c(E)$ is the critical distance  for  a perfectly rigid lattice (which decreases as $E$ increases, as shown in Fig. 2.a),  $u_1(T)$ is the  1-dimensional amplitude of thermal fluctuations (we used the Debye model) which increases with $T$,  and $c$ is a dimensionless number which (for several crystals and propagating ions, different than the ones we study) was found through data and simulations to be $1< c< 2$.  At large enough energies $\rho_c (E,T) \simeq c~u_1(T)$ and thus  as $T$ increases the strings are planes become thicker, the channels narrower and $\psi_c$ smaller (see Fig. 2). Using this formalism  with $c=2$ we could reproduce data on channeling angles of B and P ions in Si measured at room $T$ provided by Hobler~\cite{Hobler} (shown in green for B in Fig.1.b). 
As shown in Figs. 1.b  and 2.b the critical angle $\psi_c$ increases with decreasing energy until the critical distance for channeling  $\rho_c$ approaches  the radius of the channel, at which point  $\psi_c$ goes zero (see Fig. 1.a). At still lower energies, $\rho_c$ should be larger than the radius of the channel,  what indicates that channeling is  not possible (thus $\psi_c=0$). 

\begin{figure} 
\includegraphics[width=.50\textwidth]{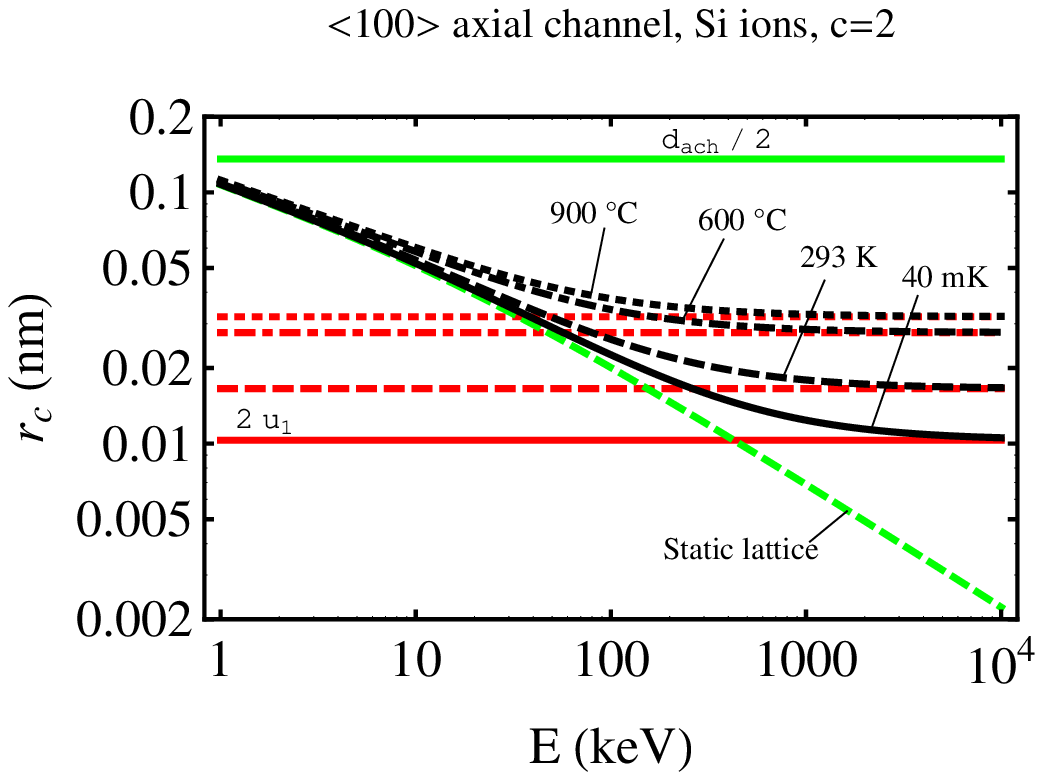} 
\includegraphics[width=.47\textwidth]{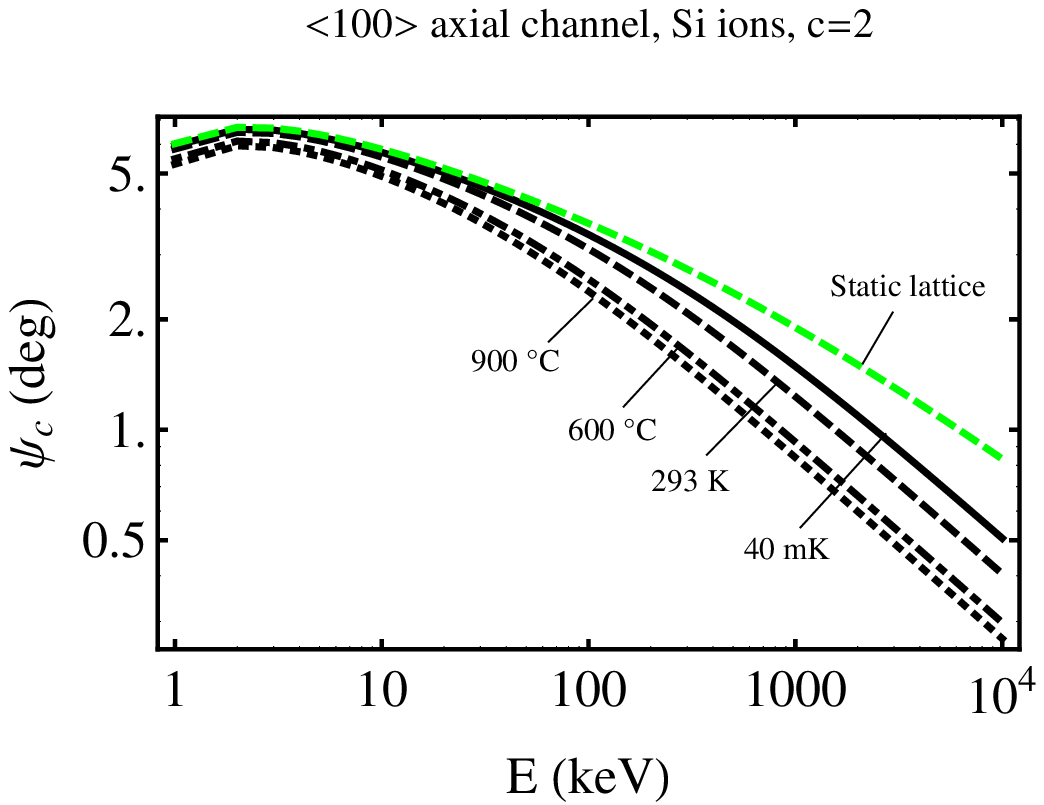} 
\caption{a. (left) Critical channeling distance $r_c$ and b. (right) critical channeling angles $\psi_c$  for a Si ion propagating in the $<$100$>$ channel of a Si crystal, for a static lattice (dashed green) and $T$ corrected (solid black lines) at different $T$ with $c=2$ (2$u_1$  shown in red and the radius of the channel in solid green)} 
\label{fig 2} 
\end{figure} 
\begin{figure} 
\includegraphics[width=.45\textwidth]{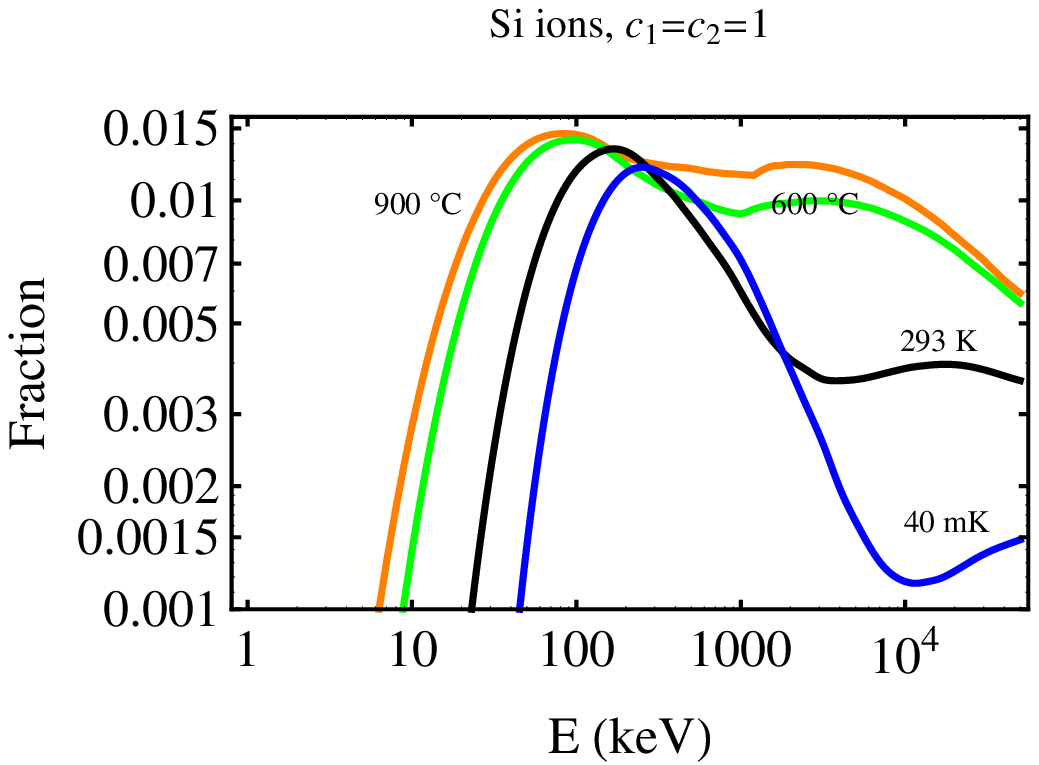} 
\includegraphics[width=.45\textwidth]{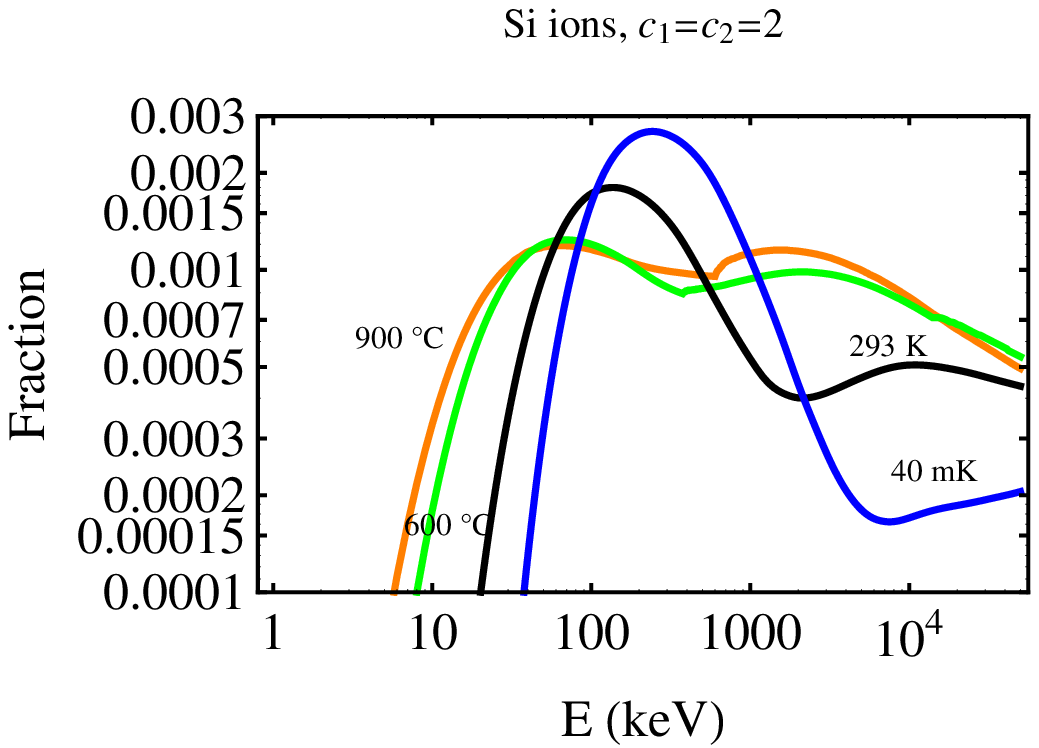} 
\caption{Channeling fractions for a Si ions ejected from lattice sites in a Si crystal  as function of the ion energy for different temperatures $T$ and $T$ effects computed with a. (left) $c=1$ and b. (right) $c=2$.} 
\label{fig 3} 
\end{figure} 
\begin{figure} 
\includegraphics[width=.45\textwidth]{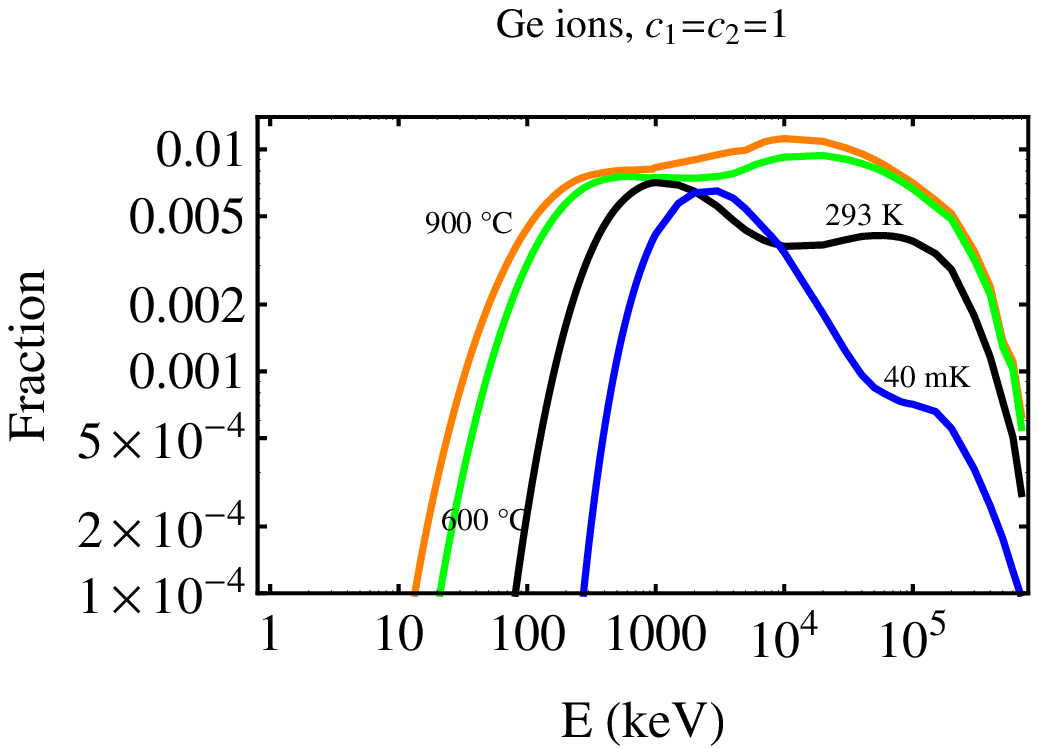} 
\includegraphics[width=.45\textwidth]{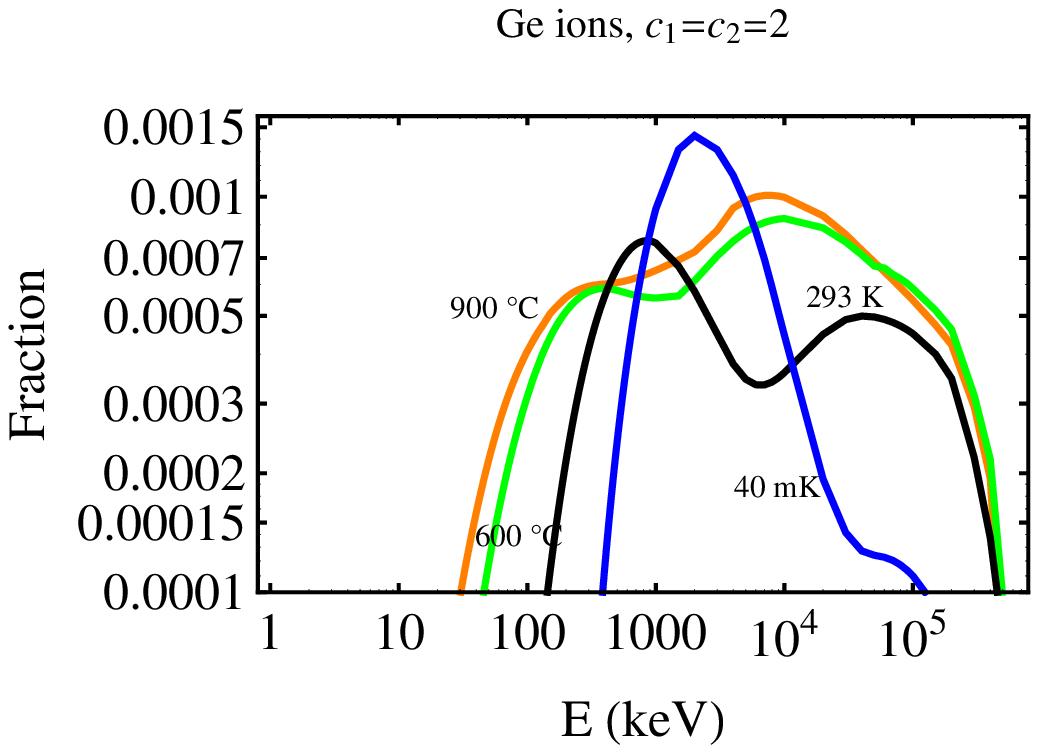} 
\caption{Same as Fig. 3 but for Ge ions ejected from lattice sites in a Ge crystal.} 
\label{fig 4} 
\end{figure} 

The channeling of an ion depends not only on the initial angle its trajectory makes with a string or plane, but also on its initial position. The DAMA collaboration~\cite{DAMA}  calculated the channeling fraction for ions starting their motion close to the middle of the channel, where channeling happens if $\psi <\psi_c$.
However this is not the case in dark matter detectors, since the recoiling ions start their motion at or close to their original lattice sites (and leave those sites empty). For recoiling  ions, blocking effects (neglected in the DAMA calculation) are important. In fact in a perfect lattice no recoiling ion would be channeled (because of what Lindhard called the ``rule of reversibility"), but due to lattice vibrations the collision with a WIMP may happen while  an atom is displaced with respect to the string or plane where it belongs. If it is initially far enough it may be channeled.
\begin{figure} 
\includegraphics[width=.45\textwidth]{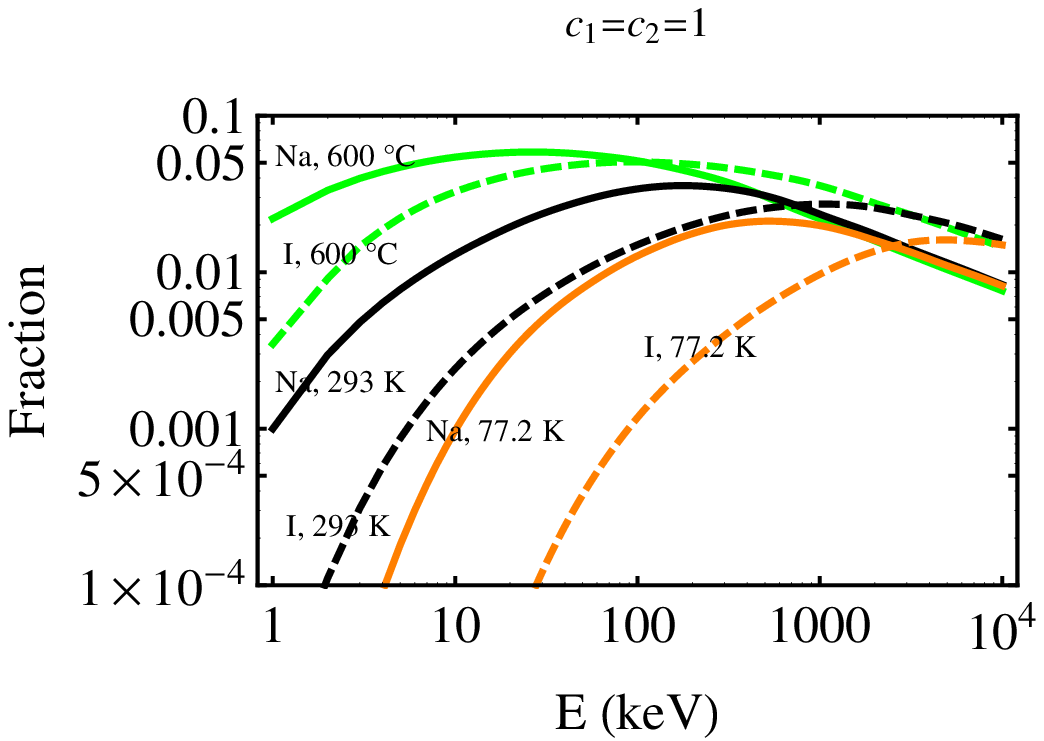} 
\includegraphics[width=.55\textwidth]{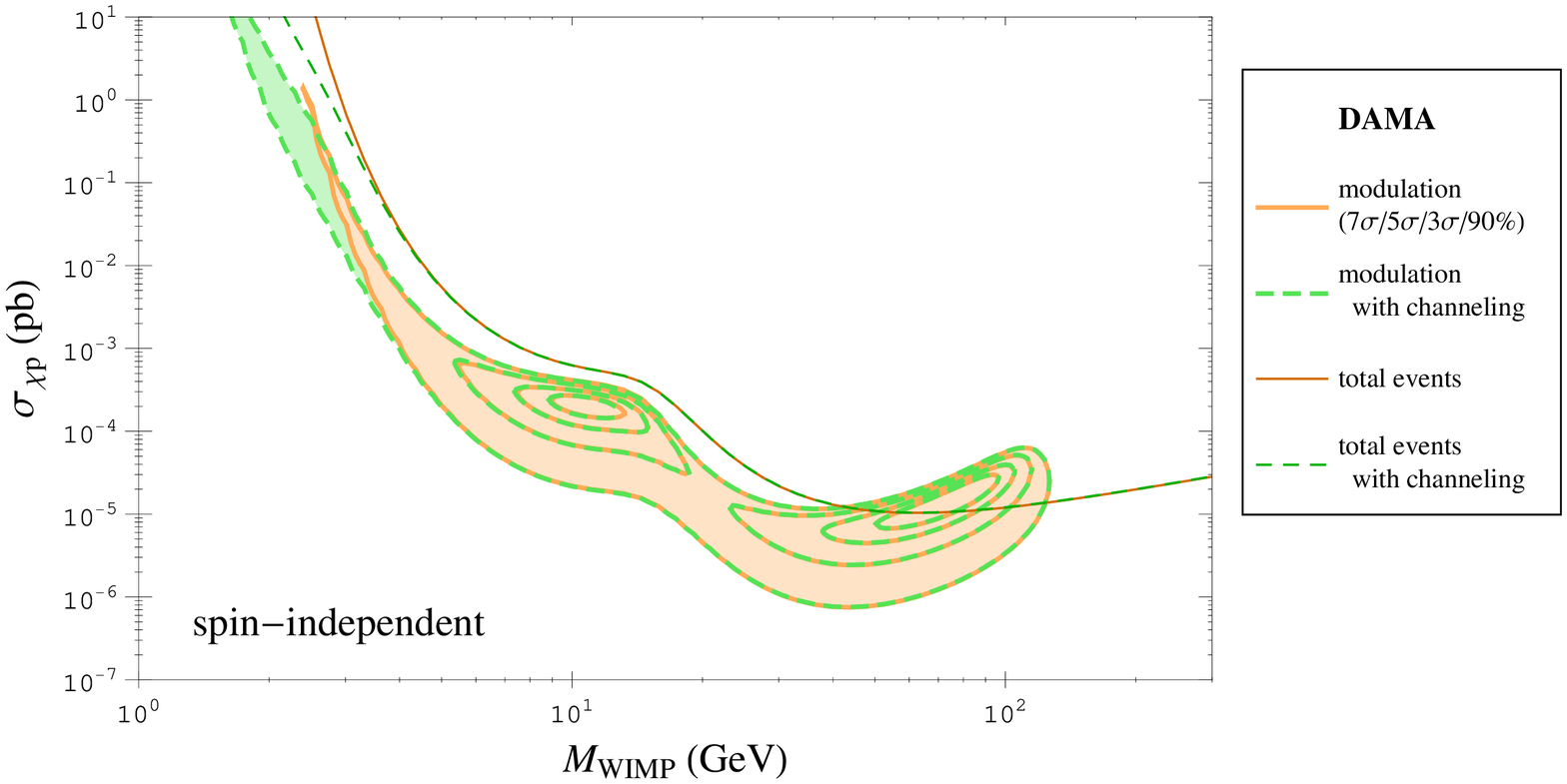} 
\caption{a. (left) Upper bounds to the channeling fraction of Na and I  ions ejected from lattice sites  in a NaI crystal for several temperatures (with $c=1$). b. (right) Using the T= 293 K curve as the channeling fraction in the DAMA detectors the region of compatibility with the DAMA signal for WIMPs with spin-independent interaction is the same with and without channeling at less than 7$\sigma$~\cite{Savage:2010tg}.} 
\label{fig4} 
\end{figure} 
For an axial channel, the probability distribution function of the perpendicular distance to the row of the colliding atom due to thermal vibrations can be represented by a Gaussian, $g(\rho)=(\rho/u_1^2) e^{(-\rho^2/2u_1^2)}$. Thus, in our model the probability  that an ion is channeled
is given by the fraction of nuclei which can be found at a distance larger than a minimum distance $\rho_{i,\rm min}$ (determined by $\rho_c$ and the initial recoil angle) from the string at the moment of collision, $P_{Ch}= \int_{\rho_{i,\rm min}}^{\infty}{dr g(\rho)}=e^{(-\rho_{i,\rm min}^2/2u_1^2)}$. Notice that any uncertainty in $\rho_c$ is exponentially enhanced in the channeling fraction.  Similar equations apply to planar channels. To obtain the total geometrical channeling fraction, we average the channeling probability  over initial recoil  directions, assuming they are isotropically distributed.  We do it numerically by performing a Riemann sum once the sphere of directions has been divided using a Hierarchical Equal Area iso-Latitude Pixelization (HEALPix)~\cite{HEALPix:2005} (a novel use for HEALPix). 
 
 The channeling fractions we obtain are strongly $T$ dependent (see e.g. Figs. 2,3 and 4.a). As $T$ increases the probability of finding atoms far from their equilibrium lattice sites increases, which increases the channeling fractions, but the critical distances $\rho_c$ become larger ($\simeq c u_1$ at large enough energies) which decreases the channeling fractions.  The fractions are smaller for larger values of $c$, as can be seen for Si and Ge in Fig. 3 and 4.  We expect the fractions to be in between those for $c=1$ and those for $c=2$.  Moreover, we have not included any dechanneling effects due to impurities or dopants. So the fractions we produced must be considered upper bounds.
 This is even more so for NaI and CsI,  for which we have not found data equivalent to those provided by Hobler for Si (useful for Ge too, which has the same structure of Si) so we can provide upper bounds on the channeling fractions instead of estimates, even forgetting about dechanneling effects.  Upper bounds on channeling fractions in NaI are shown in Fig. 4.a for $c=1$ and several temperatures. Using the fractions in Fig. 5.a corresponding to the temperature of the DAMA/NaI and the DAMA/LIBRA experiments, T= 293 K, the difference in the WIMP region corresponding to the DAMA annual modulation signal differs only at 7$\sigma$ is channeling is included or not included (as can be seen in Fig. 5.b taken from Ref.~\cite{Savage:2010tg}).
 
 To conclude, the effect of blocking is  important to understand the channeling of ions ejected from lattice sites by interactions with dark matter particles. Analytic  models give good qualitative results but  channeling data and or simulations  are necessary to get reliable quantitative results. Finally, the daily modulation of the scintillation or ionization in crystalline detectors due to channeling, a potentially background free dark matter signal, deserves further investigation.

\end{document}